\documentclass[journal]{IEEEtran}

\ifCLASSINFOpdf
\else
   \usepackage[dvips]{graphicx}
\fi
\usepackage{url}

\usepackage{graphicx, multirow}
\usepackage{booktabs}
\usepackage{amsmath,amssymb,amsfonts}
\usepackage{bbding}

\allowdisplaybreaks[4]

\begin{document}

\title{UniArray: Unified Spectral-Spatial Modeling for Array-Geometry-Agnostic Speech Separation}

\author{Weiguang Chen, Junjie Zhang, Jielong Yang, Eng Siong Chng, \IEEEmembership{Senior Member, IEEE}, and Xionghu Zhong, \IEEEmembership{Member, IEEE}
% \thanks{This paragraph of the first footnote will contain the date on which you submitted your paper for review. It will also contain support information, including sponsor and financial support acknowledgment. For example, ``This work was supported in part by the U.S. Department of Commerce under Grant BS123456.'' }
\thanks{W. Chen, J. Zhang, and X. Zhong are with the College of Computer Science and Electronic Engineering, Hunan University, Hunan, China. (emails: \{cwg,~zjj,~and xzhong\}@hnu.edu.cn).}
\thanks{We thank Dr. Yiyang Chen for his contributions to experiment design and data preprocessing.}
\thanks{J. Yang is with the School of Internet of Things Engineering, Jiangnan University, China. (email: jyang022@e.ntu.edu.sg).}
\thanks{E. S. Chng is with the School of Computer Science and Engineering, Nanyang Technological University, Singapore. (email: aseschng@ntu.edu.sg).}
% \thanks{The next few paragraphs should contain the authors' current affiliations, including current address and e-mail. For example, F. A. Author is with the National Institute of Standards and Technology, Boulder, CO 80305 USA (e-mail: author@boulder.nist.gov).}
% \thanks{S. B. Author, Jr., was with Rice University, Houston, TX 77005 USA. He is now with the Department of Physics, Colorado State University, Fort Collins, CO 80523 USA (e-mail: author@lamar.colostate.edu).}
}

% \markboth{Journal of \LaTeX\ Class Files, Vol. 14, No. 8, August 2015}
% {Shell \MakeLowercase{\textit{et al.}}: Bare Demo of IEEEtran.cls for IEEE Journals}
\maketitle

\begin{abstract}
Array-geometry-agnostic speech separation (AGA-SS) aims to develop an effective separation method regardless of the microphone array geometry.
Conventional methods rely on permutation-free operations, such as summation or attention mechanisms, to capture spatial information.
However, these approaches often incur high computational costs or disrupt the effective use of spatial information during intra- and inter-channel interactions, leading to suboptimal performance.
To address these issues, we propose UniArray, a novel approach that abandons the conventional interleaving manner.
UniArray consists of three key components: a virtual microphone estimation (VME) module, a feature extraction and fusion module, and a hierarchical dual-path separator.
The VME ensures robust performance across arrays with varying channel numbers.
The feature extraction and fusion module leverages a spectral feature extraction module and a spatial dictionary learning (SDL) module to extract and fuse frequency-bin-level features, allowing the separator to focus on using the fused features.
The hierarchical dual-path separator models feature dependencies along the time and frequency axes while maintaining computational efficiency.
Experimental results show that UniArray outperforms state-of-the-art methods in SI-SDRi, WB-PESQ, NB-PESQ, and STOI across both seen and unseen array geometries. 
\end{abstract}

\begin{IEEEkeywords}
array-geometry-agnostic, multi-channel, speech separation, spatial dictionary, virtual microphone.
% Enter key words or phrases in alphabetical order, separated by commas. For a list of suggested keywords, send a blank e-mail to keywords@ieee.org or visit \url{http://www.ieee.org/organizations/pubs/ani_prod/keywrd98.txt}
\end{IEEEkeywords}

\IEEEpeerreviewmaketitle

\section{Introduction}

\IEEEPARstart{S}{peech} separation aims to recover individual speech signals from mixtures, enabling applications such as automatic speech recognition~\cite{yu2020audio}, speaker diarization~\cite{bullock2020overlap}, and identification~\cite{nagrani2017voxceleb}.
Recent advances in self-attention~\cite{subakan2021attention, yang2022tfpsnet, wang2023tf} and state-space models~\cite{chen23g_interspeech, jiang2024dual} have significantly improved separation performance. Multi-channel techniques leveraging microphone arrays further enhance separation by incorporating spatial features~\cite{gu2020enhancing, quan2022multichannel, AroudiUF22, quan2024spatialnet}.

However, existing multi-channel methods are typically tailored to fixed array geometries (e.g., linear or circular) and a predefined number of channels~\cite{quan2022multichannel, AroudiUF22, quan2024spatialnet, quan2022multi}. This reliance on static configurations restricts their applicability in real-world scenarios where array geometries and channel numbers vary.
To reduce the time and effort required for re-training and facilitate array-geometry-agnostic speech separation (AGA-SS), VarArray~\cite{yoshioka2022vararray} addressed this issue by introducing the transform-average-concatenate (TAC) module~\cite{luo2020end}, which facilitates channel-invariant processing by averaging features across channels. While effective for handling varying channel numbers, the averaging operation neglects inter-channel relationships, limiting the exploitation of spatial features.
To address this limitation, attention mechanisms~\cite{wang2020neural, jukic2023flexible} and advanced triple-path architectures~\cite{xiang2022distributed, pandey2022tparn, wang23na_interspeech} have been proposed, combining intra-channel dual-path processing~\cite{luo2020dual} with inter-channel modeling. However, these methods rely on iterative interleaving of intra- and inter-channel processes, often losing critical spatial dependencies before inter-channel processing. Moreover, the added complexity of triple-path approaches significantly increases computational overhead, hindering their practicality.

In this letter, we propose UniArray, which abandons the conventional interleaving of intra- and inter-channel processes.
Instead, UniArray explicitly models spectral and spatial features at the frequency-bin level and consists of three key components: a virtual microphone estimation (VME) module, a feature extraction and fusion module, and a hierarchical dual-path separator.
The VME generates virtual microphone signals to handle varying channel numbers and ensure consistent performance across different array configurations.
The feature extraction and fusion module combines a spectral feature extraction module and a spatial dictionary learning (SDL) module to capture and integrate frequency-bin-level features.
The hierarchical dual-path separator then models dependencies along the time and frequency axes, with a patching mechanism that reduces computational complexity by aggregating adjacent time-frequency bins.
Experimental results show that UniArray outperforms existing AGA-SS methods, with average SI-SDRi improvements of $24.9\%$ and $24.3\%$ over the triple-path method across seen and unseen array geometries.

\vspace{-5pt}
\section{UniArray}

\begin{figure*}[htbp]
\centerline{\includegraphics[width=0.85\linewidth]{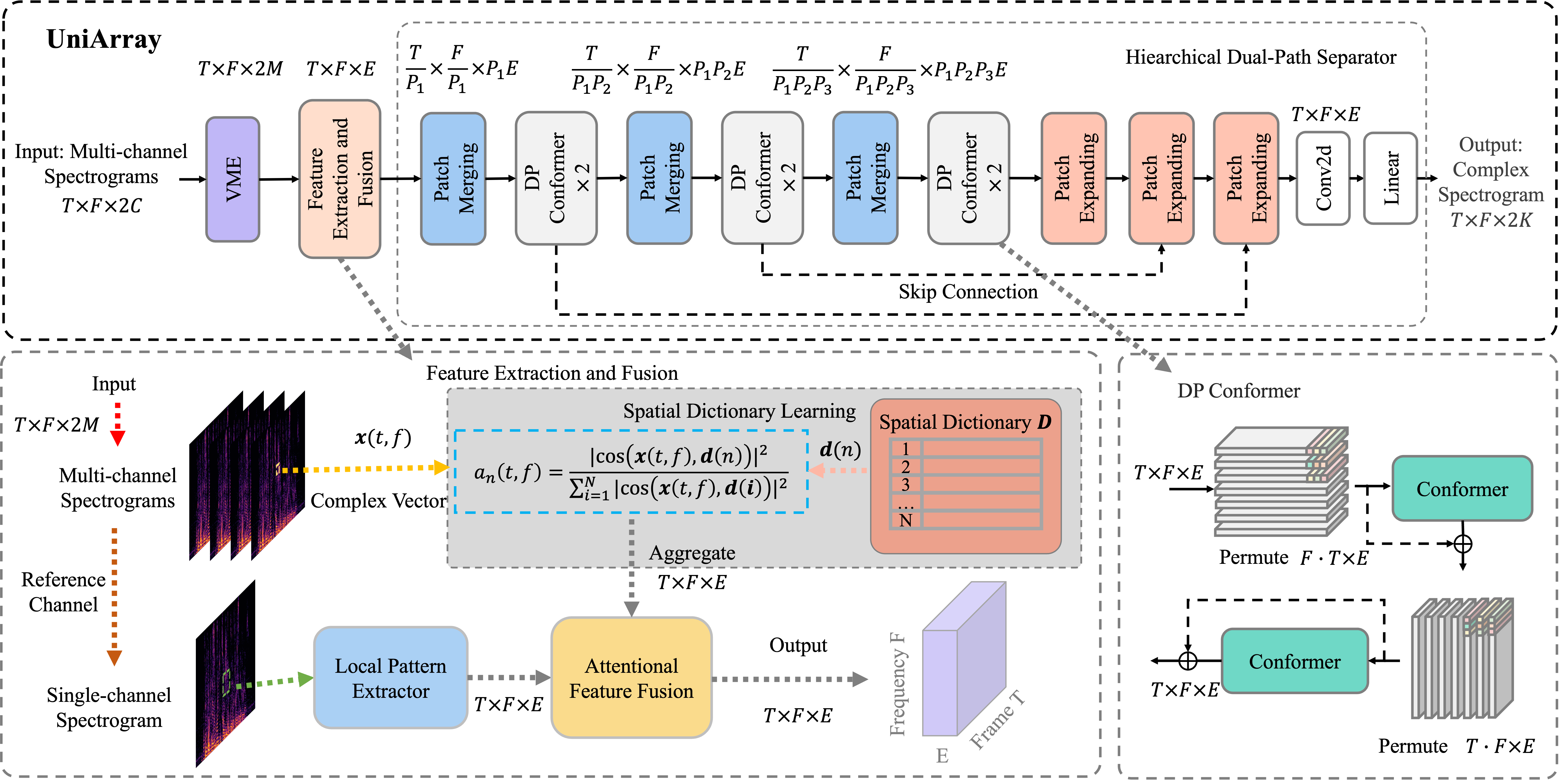}}
\vspace{-10pt}
\caption{The overall architecture of the proposed UniArray consists of three key modules: VME, feature extraction and fusion, and a hierarchical dual-path separator. The VME module generates virtual microphone signals to augment the number of channels up to the maximum $M$. The feature extraction and fusion module captures both spectral and spatial features at the frequency-bin level. Finally, the hierarchical dual-path separator estimates the clean spectrogram for each speaker.}
\label{fig:archi}
\vspace{-20pt}
\end{figure*}

The overall architecture of the proposed UniArray is shown in Fig.~\ref{fig:archi}. It comprises three main components: a virtual microphone estimation (VME) module, a feature extraction and fusion module, and a hierarchical dual-path separator. As a frequency-domain approach, UniArray processes multi-channel complex spectrograms with a shape of $T \times F \times 2C$ ($C \leq M$), where $T$ is the frame length, $F$ is the number of frequency bins, $C$ is the input channel number, and $M$ is the maximum channel number.
The VME module generates virtual microphone signals until the channel dimension reaches $M$.
The feature extraction and fusion module extracts spectral and spatial information at the frequency-bin level and fuses them for the following separator.
Then the separator reconstructs the clean complex spectrogram for each speaker, with a shape of $T \times F \times 2K$, where $K$ denotes the number of speakers.

\vspace{-5pt}
\subsection{Virtual Microphone Estimation}

Virtual microphone estimation (VME) is a technique to enhance spatial resolution by generating virtual microphone signals, widely studied in beamforming~\cite{katahira2016nonlinear, yamaoka2019cnn, ochiai2021neural, segawa2024neural}. By interpolating between real microphone signals, VME produces spatially coherent virtual signals, effectively improving spatial resolution while preserving the integrity of the original audio.
In this study, we explore VME for the first time in the context of the AGA-SS task. The adopted interpolation-based VME method~\cite{katahira2016nonlinear} generates the phase and amplitude of virtual signals by interpolating between two real microphones.
Unlike previous works~\cite{katahira2016nonlinear, yamaoka2019cnn, ochiai2021neural, segawa2024neural}, which focus on a fixed pair of real microphones, the AGA-SS task involves variable channel numbers. To address this, we estimate virtual microphones until the maximum channel number $M$ is reached. Virtual microphones are distributed evenly across real microphone pairs as follows: 
\vspace{-4pt}
\begin{align}
    n_i &= \lfloor \frac{N_v}{N_p} \rfloor + \delta(i), \\
    \delta(i) &= \left\{
    \begin{aligned}
    & 1,\quad \text{if} \ i \ \leq N_v \ \text{mod} \ N_p \\
    & 0,\quad \text{if} \ i \ \textgreater \ N_v \ \text{mod} \ N_p
    \end{aligned}
    \right.
\end{align}
where $n_i$ denotes the virtual microphone numbers of pair $i$,
$N_v$ and $N_p$ represent the number of virtual microphones and real microphone pairs, respectively, and $\delta(i)$ ensures the first $N_v \ \text{mod} \ N_p$ pairs receive one extra virtual microphone. Adjacent microphones are paired, with the first and last microphones forming a pair. Thus, the number of real microphone pairs $N_p$ equals the number of input microphones $C$.

\begin{table*}[htbp]
\centering
\vspace{-10pt}
\caption{Performance comparison with baselines for seen microphone arrays during training. Due to space restrictions, only 8/6/4/2 microphones are reported.}
\vspace{-5pt}
\label{table:result_seen}
\resizebox{\linewidth}{!}{
\begin{tabular}{lcccc|cccc|cccc|cccc}
\toprule
\multicolumn{1}{c}{\multirow{2}{*}{\textbf{Method}}} & \multicolumn{4}{c}{\textbf{C-8-5}}                                                                                                                                                                       & \multicolumn{4}{c}{\textbf{C-6-5}}                                                                                                                                                                       & \multicolumn{4}{c}{\textbf{C-4-5}}                                                                                                                                                                       & \multicolumn{4}{c}{\textbf{L-2-10}}                                                                                                                                                                      \\ \cmidrule{2-17} 
\multicolumn{1}{c}{}                                 & \textbf{\begin{tabular}[c]{@{}c@{}}WB-\\ PESQ\end{tabular}} & \textbf{\begin{tabular}[c]{@{}c@{}}NB-\\ PESQ\end{tabular}} & \textbf{STOI}  & \textbf{\begin{tabular}[c]{@{}c@{}}SI-\\ SDRi\end{tabular}} & \textbf{\begin{tabular}[c]{@{}c@{}}WB-\\ PESQ\end{tabular}} & \textbf{\begin{tabular}[c]{@{}c@{}}NB-\\ PESQ\end{tabular}} & \textbf{STOI}  & \textbf{\begin{tabular}[c]{@{}c@{}}SI-\\ SDRi\end{tabular}} & \textbf{\begin{tabular}[c]{@{}c@{}}WB-\\ PESQ\end{tabular}} & \textbf{\begin{tabular}[c]{@{}c@{}}NB-\\ PESQ\end{tabular}} & \textbf{STOI}  & \textbf{\begin{tabular}[c]{@{}c@{}}SI-\\ SDRi\end{tabular}} & \textbf{\begin{tabular}[c]{@{}c@{}}WB-\\ PESQ\end{tabular}} & \textbf{\begin{tabular}[c]{@{}c@{}}NB-\\ PESQ\end{tabular}} & \textbf{STOI}  & \textbf{\begin{tabular}[c]{@{}c@{}}SI-\\ SDRi\end{tabular}} \\ \midrule
\textbf{FasNet-TAC}~\cite{luo2020end}                                           & 1.89                                                        & 2.46                                                        & 75.96          & 10.03                                                       & 1.87                                                        & 2.45                                                        & 75.72          & 9.99                                                        & 1.86                                                        & 2.43                                                        & 75.49          & 9.92                                                        & 1.77                                                        & 2.32                                                        & 74.23          & 9.14                                                        \\
\textbf{VarArray}~\cite{yoshioka2022vararray}                                             & 1.56                                                        & 1.96                                                        & 51.56          & 9.68                                                        & 1.55                                                        & 1.96                                                        & 51.72          & 9.62                                                        & 1.55                                                        & 1.96                                                        & 51.59          & 9.60                                                        & 1.52                                                        & 1.93                                                        & 52.88          & 9.03                                                        \\
\textbf{VarArray+CA}~\cite{jukic2023flexible}                                          & 1.59                                                        & 1.98                                                        & 56.64          & 9.56                                                        & 1.58                                                        & 1.98                                                        & 56.92          & 9.51                                                        & 1.58                                                        & 1.98                                                        & 56.97          & 9.47                                                        & 1.54                                                        & 1.92                                                        & 57.94          & 8.87                                                        \\
\textbf{SDNet}~\cite{wang23na_interspeech}                                                & 2.25                                                        & 2.99                                                        & 83.01          & 11.12                                                       & 2.24                                                        & 2.98                                                        & 82.94          & 11.10                                                       & 2.23                                                        & 2.96                                                        & 82.58          & 11.08                                                       & 2.15                                                        & 2.84                                                        & 81.41          & 10.56                                                       \\ \midrule
\textbf{UniArray}                                    &                                                             &                                                             &                &                                                             &                                                             &                                                             &                &                                                             &                                                             &                                                             &                &                                                             &                                                             &                                                             &                &                                                             \\
\textbf{\quad +SDL}                                        & \textbf{2.96}                                               & \textbf{3.58}                                               & \textbf{90.67} & \textbf{13.76}                                              & \textbf{2.96}                                               & \textbf{3.58}                                               & \textbf{90.67} & \textbf{13.76}                                              & \textbf{2.96}                                               & \textbf{3.58}                                               & \textbf{90.66} & \textbf{13.75}                                              & \textbf{2.89}                                               & \textbf{3.52}                                               & 90.14          & \textbf{13.50}                                              \\
\textbf{\quad +FSDL}                                       & 2.94                                                        & 3.57                                                        & 90.61          & 13.62                                                       & 2.94                                                        & 3.57                                                        & 90.62          & 13.61                                                       & 2.94                                                        & 3.57                                                        & 90.55          & 13.58                                                       & \textbf{2.89}                                               & \textbf{3.52}                                               & \textbf{90.18} & 13.41                                                       \\ \bottomrule
\end{tabular}
}
\vspace{-20pt}
\end{table*}

\vspace{-10pt}
\subsection{Feature Extraction and Fusion}
\label{subsec:feat}

For the AGA-SS task, effectively leveraging spatial information alongside spectral features is essential. To address this, we propose a unified approach that jointly models spectral and spatial patterns. As illustrated in Fig.~\ref{fig:archi}, the method employs a local spectro-temporal pattern extractor for spectral features, a spatial dictionary learning (SDL) module for spatial cues, and an attentional feature fusion module to integrate them. Note that the dimensionality of the spectral, spatial, and fused features is unified and denoted as $E$.

\noindent\textbf{Local Pattern Extractor.}
Building on our previous work~\cite{xiong22_interspeech}, which demonstrated the effectiveness of local spectro-temporal features in speech enhancement, we utilize a 2-D convolutional neural network to extract these patterns from the reference channel ($\text{default} = 0$) to enhance speech quality.

\noindent\textbf{Spatial Dictionary Learning (SDL).}
The SDL module captures spatial cues from multi-channel spectrograms by modeling both phase and magnitude differences at the frequency-bin level. The complex vector at frame $t$ and frequency $f$ can be represented as $\boldsymbol x(t,f) = [x_1(t,f), x_2(t,f), \cdots, x_M(t,f)]^\top$.
To model spatial features, a learnable spatial dictionary $\boldsymbol{D} = [\boldsymbol{d}(1), \cdots, \boldsymbol{d}(n), \cdots, \boldsymbol{d}(N)] \in \mathbb C^{M \times N}$ is constructed.
This dictionary projects complex vectors onto a hypersphere, generating spatial embeddings $\boldsymbol{a}(t,f) = [a_1(t,f), \dots, a_n(t,f), \dots, a_N(t,f)]$, as follows
\begin{align}
    a_{n}^\prime(t,f) &= \left| \frac{\boldsymbol{d}^\top(n) \boldsymbol{x}(t,f)}{| \boldsymbol{d}^\top(n)| |\boldsymbol{x}(t,f)|} \right|^2, \\
    x_{avg}(t,f) &= \frac{1}{M} \sum_{i}^{M} | x_i(t,f) |, \\
    a_{n}(t,f) &= x_{avg}(t,f) \times a_{n}^\prime(t,f).
\end{align}
where $\left|\cdot\right|$ represents the Euclidean norm. To make the SDL module focusing on learning spatial features, both the complex vectors and the dictionary items are normalized.
To prioritize frequency bins with significant power, the spatial embeddings are scaled by the average magnitude spectrogram.
Following our previous work~\cite{chen21t_interspeech, chen2024enhancing}, the SDL module is shared across different frequency bands. However, considering that spatial cues for the same speaker can vary across frequency bands, this study also explores a frequency-dependent SDL (FSDL) for comparison.

\noindent \textbf{Attentional Feature Fusion.}
To integrate spectral and spatial features, we adopt an iterative attentional feature fusion mechanism~\cite{dai2021attentional}, which aggregates local and global contexts and assigns attention scores to each feature map. The final fused feature map $\mathbf{X} \in \mathbb{R}^{T \times F \times E}$ is obtained as the weighted sum of the spectral and spatial features.
% To complement spectral and spatial features, we introduce an iterative attentional feature fusion scheme~\cite{dai2021attentional} to fuse the spatial and spectral features.
% It aggregates local and global contexts iteratively and assigns attention scores to each feature map.
% Finally, the weighted sum of two features is taken as the fused feature map $\mathbf{X} \in \mathbb R^{T \times F \times E}$.

\vspace{-10pt}
\subsection{Hierarchical Dual-path Separator}

As introduced in Sec.~\ref{subsec:feat}, both the local pattern extractor and SDL modules operate at the time-frequency bin level. For each frequency bin, its relationships span both other frequency bins in the same frame and other frames in the same frequency band. To capture these dependencies, we propose a dual-path separator that models time and frequency dependencies, differing from traditional dual-path methods that focus solely on intra- and inter-chunk processes along the time axis~\cite{yi2020dprnn, chen2020dual, jiang2024dual, ravenscroft2024combining}. However, directly processing the output tensor at its original resolution is computationally expensive. To address this, we introduce a hierarchical dual-path separator for efficiency.

As shown in Fig.~\ref{fig:archi}, this approach progressively groups adjacent time-frequency bins into larger units using patch merging layers with a window size of $P_i \times P_i$. Each merging layer includes a convolutional operation followed by layer normalization. After each merging step, a dual-path Conformer module is applied to capture feature dependencies along both time and frequency axes. Specifically, the module consists of two Conformer layers: the first models temporal dependencies within each frequency band, while the second captures correlations across frequency bands within each time frame. A symmetric deconvolutional decoder then restores the feature map to its original resolution using patch expanding layers.

Unlike methods such as~\cite{yang2022tfpsnet, wang2023tf}, which introduce a third path at the cost of increased computational complexity, UniArray focuses on efficiently modeling time and frequency dependencies. While similar to~\cite{wang2023dasformer}, UniArray adopts a hierarchical dual-path design, significantly reducing computational overhead and improving practicality for real-world applications.

\vspace{-5pt}
\section{Experiments and Results}

\subsection{Experimental Setup}

\noindent \textbf{Dataset.}
Following~\cite{quan2022multi}, we generate a spatialized version of the WSJ0-2mix dataset~\cite{hershey2016deep} to evaluate our proposed method. An 8-channel circular microphone array with a radius of 5 cm is simulated at a 16 kHz sampling rate. The reverberation time ranges from 0.1 s to 1.0 s, and the overlapping speech ratio varies between 10\% and 100\%. Multi-channel diffuse noises are generated using ambient noises from the Reverb Challenge dataset~\cite{kinoshita2013reverb} and the toolkit in~\cite{habets2008generating}, with the SNR between the reverberant mixture and noise randomly sampled between 10 and 20 dB.
During training, additional geometries are derived from the 8-channel circular array by selecting subsets of microphones with indices $\{0,4\}$, $\{0,3,5\}$, $\{0,2,4,6\}$ and $\{1,2,3,5,6,7\}$. Microphone channels are permuted to encourage the model to learn spatial features across diverse array geometries.
For testing, we evaluate on both seen and unseen array geometries with varying numbers of channels. For instance, \textbf{C-8-10}'' represents a circular array with 8 channels and a radius of 10 cm, while \textbf{L-2-5}'' denotes a linear array with 2 channels and 5 cm spacing.

\noindent \textbf{Configurations}.
In our method, the short-time Fourier transform (STFT) uses a 32 ms Hanning window with a 16 ms shift. The kernel and stride sizes of the local pattern extractor are 5 and 1, respectively. Both the spatial dictionary size $N$ and spectral embedding size $E$ are set to 64.
The window sizes for the three patch merging layers are configured as $[1 \times 1, 2 \times 2, 2 \times 2]$. For the Conformer module, we adopt the same configuration as in~\cite{quan2022multichannel}.
Model training uses the scale-invariant signal-to-distortion ratio (SI-SDR)\cite{le2019sdr} loss with utterance-level permutation invariant training (uPIT)\cite{kolbaek2017multitalker}, targeting clean reverberant speech. All models, including baselines and our method, are trained for 100 epochs.

\begin{table*}[htbp]
\centering
\vspace{-10pt}
\caption{Performance comparison with baselines for unseen microphone arrays during training in terms of SI-SDRi.}
\vspace{-5pt}
\label{table:result_unseen}
\resizebox{\linewidth}{!}{
    \begin{tabular}{lccccccccccc|cc}
    \toprule
    \multicolumn{1}{c}{\textbf{Method}} & \textbf{C-8-10} & \textbf{C-8-4.25} & \textbf{C-7-4.25} & \textbf{C-6-4.25} & \textbf{C-5-4.25} & \textbf{C-4-4.25} & \textbf{C-4-3} & \textbf{C-3-3} & \textbf{L-3-3} & \textbf{L-2-5} & \textbf{Avg} & \textbf{Para. (M)} & \textbf{MACs (G/s)}   \\ \midrule
    \textbf{FasNet-TAC}~\cite{luo2020end}                 & 9.64            & 9.97              & 10.01             & 9.93              & 9.90              & 9.85              & 9.72           & 9.62           & 9.01           & 8.99           & 9.66   & 2.76 & 18.17        \\
    \textbf{VarArray}~\cite{yoshioka2022vararray}                   & 9.47            & 9.62              & 9.66              & 9.58              & 9.57              & 9.56              & 9.47           & 9.41           & 8.83           & 8.88           & 9.41   & 2.77 & 0.75        \\
    \textbf{VarArray+CA}~\cite{jukic2023flexible}                & 9.29            & 9.51              & 9.57              & 9.50              & 9.47              & 9.45              & 9.32           & 9.26           & 8.71           & 8.73           & 9.28     & 7.13 & 1.84      \\
    \textbf{SDNet}~\cite{wang23na_interspeech}                      & 11.11           & 11.03             & 11.00             & 11.01             & 10.93             & 10.95             & 10.91          & 10.83          & 10.29          & 10.27          & 10.83    & 0.89 & 22.56      \\ \midrule
    \textbf{UniArray}                   &                 &                   &                   &                   &                   &                   &                &                &                &                &                \\
    \textbf{\quad +SDL}                       & \textbf{13.63}  & \textbf{13.72}    & \textbf{13.33}    & \textbf{13.69}    & \textbf{13.28}    & \textbf{13.69}    & \textbf{13.49} & \textbf{13.41} & \textbf{13.28} & \textbf{13.09} & \textbf{13.46} & 7.20 & 12.96 \\
    \textbf{\quad +FSDL}                      & 13.28           & 13.55             & 13.30             & 13.53             & 13.24             & 13.52             & 13.18          & 13.12          & 13.14          & 12.87          & 13.27  & 7.46 & 12.96        \\ \bottomrule
    \end{tabular}
}
\vspace{-15pt}
\end{table*}

\begin{table}[htbp]
\centering
% \vspace{-5pt}
\caption{Ablation study of VME and SDL modules in terms of SI-SDRi. ``\Checkmark'' denotes the existence of corresponding module and ``\XSolidBrush'' indicate the opposite.}
\vspace{-5pt}
\label{table:result_ablation}
\resizebox{\linewidth}{!}{
\begin{tabular}{@{}lcccccc@{}}
\toprule
\multicolumn{1}{c}{\multirow{2}{*}{\textbf{Method}}} & \multirow{2}{*}{\textbf{VME}} & \multirow{2}{*}{\textbf{SDL}} & \multirow{2}{*}{\textbf{\begin{tabular}[c]{@{}c@{}}Seen\\ Geom\end{tabular}}} & \multicolumn{3}{c}{\textbf{Unseen Geom}}                                                                                                                                                                                  \\ \cmidrule{5-7} 
\multicolumn{1}{c}{}                                 &                               &                               &                                                                               & \textbf{\begin{tabular}[c]{@{}c@{}c@{}}Seen\\ \#Ch\\ \{2,3,4,6,8\}\end{tabular}} & \textbf{\begin{tabular}[c]{@{}c@{}c@{}}Unseen\\ \#Ch\\ \{5,7\}\end{tabular}} & \textbf{\begin{tabular}[c]{@{}c@{}c@{}}Unseen\\ \#Ch\\ \{1\}\end{tabular}} \\ \midrule
\textbf{FasNet-TAC}~\cite{luo2020end}                                  & -                             & -                             & 9.76                                                                          & 9.59                                                                        & 9.96                                                                  & 7.35                                                                \\
\textbf{VarArray}~\cite{yoshioka2022vararray}                                    & -                             & -                             & 9.13                                                                          & 9.35                                                                        & 9.62                                                                  & 7.29                                                                \\
\textbf{VarArray+CA}~\cite{jukic2023flexible}                                 & -                             & -                             & 8.99                                                                          & 9.22                                                                        & 9.52                                                                  & 7.13                                                                \\
\textbf{SDNet}~\cite{wang23na_interspeech}                                       & -                             & -                             & 10.96                                                                         & 10.8                                                                        & 10.97                                                                 & 5.65                                                                \\ \midrule
\multirow{6}{*}{\textbf{UniArray}}                   & \XSolidBrush                             & \XSolidBrush                             & 13.48                                                                         & 13.27                                                                       & 12.50                                                                 & 3.40                                                                \\
                                                     & \XSolidBrush                             & SDL                             & 13.42                                                                         & 13.27                                                                       & 12.33                                                                 & -1.49                                                               \\
                                                     & \XSolidBrush                             & FSDL                             & 13.45                                                                         & 13.26                                                                       & 12.60                                                                 & 3.00                                                                \\
                                                     & \Checkmark                             & \XSolidBrush                             & 13.38                                                                         & 13.25                                                                       & 13.10                                                                 & 3.69                                                                \\
                                                     & \Checkmark                             & SDL                             & \textbf{13.69}                                                                & \textbf{13.50}                                                              & \textbf{13.31}                                                        & 4.17                                                                \\
                                                     & \Checkmark                             & FSDL                             & 13.56                                                                         & 13.27                                                                       & 13.27                                                                 & \textbf{9.87}                                                       \\ \bottomrule
\end{tabular}
}
\vspace{-15pt}
\end{table}

\noindent \textbf{Evaluation metric.}
In addition to SI-SDR improvement (SI-SDRi), we assess performance using additional metrics for speech quality evaluation, including narrow-band PESQ (NB-PESQ)\cite{rix2001perceptual}, wide-band PESQ (WB-PESQ)\cite{rix2001perceptual}, and Short-Time Objective Intelligibility (STOI)~\cite{taal2011algorithm}.

\vspace{-10pt}
\subsection{Comparison with Baselines}

As shown in Tab.~\ref{table:result_seen}, we compare our proposed method with competitive baselines.
FasNet-TAC~\cite{luo2020end} is a time-domain method based on a dual-path RNN architecture and the TAC module, designed for microphone permutation and number invariant processing.
VarArray~\cite{yoshioka2022vararray} is a frequency-domain method using a Conformer model and TAC module, proposed for AGA-SS.
VarArray+CA~\cite{jukic2023flexible} is a viariance of VarArray where the average operation in TAC module is replaced by a channel attention module.
SDNet~\cite{wang23na_interspeech} is a recently proposed method for ad-hoc arrays, based on stream attention and dual-feature learning, and it demonstrates superiority over other ad-hoc array speech separation methods.

By comparing the experimental results on the seen microphone array during training in Tab.~\ref{table:result_seen}, we observe that: \textit{i}) Our proposed method surpasses other competitive approaches across various input channels in terms of WB-PESQ, NB-PESQ, STIO and SI-SDRi.
Particularly, UniArray achieves 13.76 SI-SDRi performance in the case of \textbf{C-8-5}, which represents a $37.2\%$ and $23.7\%$ improvement over FasNet-TAC and SDNet respectively.
Even when the input channels are reduced to 2 (i.e., \textbf{L-2-10}), UniArray still achieves a 13.5 SI-SDRi, representing a $47.7\%$ and $27.8\%$ improvement over FasNet-TAC and SDNet, respectively.
This demonstrates that UniArray is able to efficiently utilize spatial features to enhance performance.
\textit{ii}) When using an frequency-dependent SDL module (FSDL), the performance across all four metrics slightly degrades compared to the shared SDL module. This suggests that a shared SDL module is more robust and effective across different array geometries.

% When setting independent SDL module for different frequency band (i.e., FSDL), the performance in terms of four metrics are degraded slightly. This suggests that a shared SDL module is more helpful in improving speech quality.

\vspace{-5pt}
\subsection{Generalization to Unseen Microphone Arrays}

To provide a more comprehensive evaluation, we test UniArray on various unseen microphone arrays, including circular and linear arrays with different numbers of channels, radii or spacing distances.
As shown in Tab.~\ref{table:result_unseen}, UniArray generalizes well to different unseen microphone arrays, achieving performance comparable to that on the seen geometries.
Overall, UniArray with the SDL module outperforms FasNet-TAC and SDNet, with an average improvement of $39.3\%$ and $24.3\%$, respectively.
These results demonstrate that UniArray consistently improves performance across both seen and unseen array geometries.
Similar to the results on seen geometries, UniArray with the FSDL module shows slightly inferior performance compared to the SDL module.
We also compare the model size and multiply-accumulate operations (MACs) for different methods using a 4-second input duration. While UniArray has a larger model size, it achieves a smaller MACs value compared to FasNet-TAC and SDNet.

\vspace{-5pt}
\subsection{Ablation Study}

To provide a clearer understanding, we conducted an ablation study on the VME and SDL modules. The results are summarized in Tab.~\ref{table:result_ablation}, where ``\textbf{Seen Geom}'' refers to the seen geometry, and ``\textbf{\#Ch}'' refers to the number of channels. When the ``\textbf{VME}'' module is absent (\XSolidBrush), the number of channels is zero-padded to the maximum $M = 8$. When the ``\textbf{SDL}'' module is absent (\XSolidBrush), the real and imaginary parts of the multi-channel spectra are directly concatenated, resulting in an input shape of $T \times F \times 2M$.

The results demonstrate that: \textit{i})
When the input is reduced to a single channel, TAC-based methods like FasNet-TAC and VarArray generalize better than SDNet. This suggests that the TAC module enables the model to separate mixtures using spectral information even in the absence of spatial cues.
\textit{ii})
When VME and SDL modules are absent, UniArray still outperforms the baselines.
This highlights that the architecture of UniArray is superior to the interleaving of intra- and inter-channel processes for AGA-SS.
\textit{iii})
However, compared to the version with VME, UniArray without VME shows inferior performance on unseen geometries with an unseen number of channels. This suggests that the VME module is helpful for maintaining consistent separation performance in the AGA-SS task.
\textit{iv})
When VME is used, the SDL module further enhances performance through independent learning and fusion of spectral and spatial features. Additionally, the FSDL module allows UniArray to generalize well to single-channel input, with minimal impact on multi-channel performance.
In the single-channel case, $i$ and $j$ in the VME module represent the same microphone, so the virtual and real signals are identical. 
This result indicates better utilization of spectral features and offers a potential solution for unifying multi-channel AGA-SS and single-channel SS.

% \subsection{Analysis of Model Complexity}

% We analyze the model complexity of UniArray by calculating the parameter size, FLOPs, MACs and RTF.
% For a fair comparison, an 8-channel, 4-second signal is used for evaluating all methods. 
% As shown in Tab.~\ref{table:result_complexity}, UniArray has a comparable model size to VarArray+CA, though it is larger than FasNet-TAC, VarArray, and SDNet.
% In terms of MACs, UniArray has a smaller count than FasNet-TAC and SDNet, but its FLOPs are higher than those of other baselines.
% Regarding RTF, UniArray exhibits a larger value than FasNet-TAC and VarArray, but it is much smaller than SDNet.
% Notably, UniArray with the FSDL module has a lower RTF than with the SDL module.
% Overall, although UniArray's RTF is slightly above 1.0, which is higher than that of FasNet-TAC and VarArray, it remains considerably faster than SDNet and delivers better performance.

\section{Conclusion}

In this study, we propose a novel UniArray method for the AGA-SS task. UniArray leverages the VME technique to increase the number of microphones and ensure consistent performance across unseen microphone configurations. In addition to learning spectro-temporal patterns, a separate spatial dictionary learning module is designed to capture spatial information at the time-frequency bin level, which is then fused with spectral features via an attentional fusion module. A hierarchical dual-path architecture is devised to efficiently model dependencies along both the time and frequency axes.
Compared to the traditional interleaving of intra- and inter-channel processes, experimental results demonstrate that UniArray achieves superior performance across both seen and unseen geometries.

\bibliographystyle{IEEEtran}
\bibliography{mybib}

\end{document}